\documentclass[conference]{IEEEtran}
\IEEEoverridecommandlockouts

\usepackage{cite}
\def\BibTeX{{\rm B\kern-.05em{\sc i\kern-.025em b}\kern-.08em
    T\kern-.1667em\lower.7ex\hbox{E}\kern-.125emX}}

\usepackage{url}
\usepackage{booktabs}
\usepackage{amsfonts}
\usepackage{microtype}      
\usepackage{xcolor}         
\usepackage{bm}
\usepackage{bbold}
\usepackage{multirow}
\usepackage{textcomp}
\usepackage{wrapfig}
\usepackage{colortbl}
\usepackage{amsmath}
\usepackage{paralist}
\usepackage{tabto}

\definecolor{babyblueeyes}{rgb}{0.63,0.79,0.95}
\definecolor{babypink}{rgb}{0.93,0.23,0.51}
\definecolor{chromeyellow}{rgb}{1.0, 0.65, 0.0}
\definecolor{cobaltblue}{rgb}{0.16, 0.32, 0.75}
\definecolor{applegreen}{rgb}{0.55, 0.71, 0.0}
\definecolor{aqua}{rgb}{0.0, 1.0, 1.0}
\definecolor{royalblue}{rgb}{0.0, 0.14, 0.4}
\definecolor{royalazure}{rgb}{0.0, 0.22, 0.66}

\newcommand{\wn}[1]{\cellcolor{pink!60}#1}
\newcommand{\nn}[1]{\cellcolor{babypink!50}#1}
\newcommand{\strn}[1]{\cellcolor{babypink!100}#1}

\newcommand{\wpos}[1]{\cellcolor{babyblueeyes!50}#1}
\newcommand{\pp}[1]{\cellcolor{babyblueeyes!100}#1}
\newcommand{\strp}[1]{\cellcolor{royalazure!70}#1}
\newcommand{\ic}[1]{\cellcolor{gray!10}#1}

\begin{document}

\title{Objective Measurements of Voice Quality\\
}


\author{
\IEEEauthorblockN{Hira Dhamyal, Rita Singh}
\IEEEauthorblockA{\textit{Language Technologies Institute}}
\textit{Carnegie Mellon University}\\
Pittsburgh, USA \\
\{hyd, rsingh\}@cs.cmu.edu
}

\maketitle

\begin{abstract}
The quality of human voice plays an important role across various fields
like music, speech therapy, and communication, yet it lacks a
universally accepted, objective definition. Instead, voice quality is
referred to using subjective descriptors like "rough," "breathy" etc.
Despite this subjectivity, extensive research across disciplines has
linked these voice qualities to specific information about the
speaker, such as health, physiological traits, and others. Current
machine learning approaches for voice profiling rely on data-driven
analysis without fully incorporating these established correlations,
due to their qualitative nature. This paper aims to objectively
quantify voice quality by synthesizing formulaic representations from
past findings that correlate voice qualities to signal-processing
metrics. We introduce formulae for 24 voice sub-qualities based on 25
signal properties, grounded in scientific literature. These formulae
are tested against datasets with subjectively labeled voice qualities,
demonstrating their validity.
\end{abstract}


\section{Introduction}
The fact that voice carries information about the speaker has been observed and noted for centuries. In daily life, humans make myriad judgments about people based on their voice, especially the quality of their voice. Voice quality has indeed long been recognized as a fundamental aspect of human communication, influencing and supporting practice in various fields such as music, speech therapy, interpersonal communication, and speaker profiling. For example, \cite{sundberg1977acoustics} highlights the importance of vocal quality in conveying emotion and style. \cite{boone2005voice} emphasize the impact of voice quality on speech clarity and the therapeutic strategies to rehabilitate voice disorders. In interpersonal communication, the nuances of voice quality can influence perceptions of trustworthiness, attractiveness, and authority, as explored by \cite{klofstad2015perceptions}. Studies like \cite{nolan1985phonetic} discuss how vocal characteristics can be leveraged for forensic purposes, providing critical evidence in legal contexts.


Despite its significance, the assessment of voice quality has predominantly been subjective, described with terms like ``rough,'' ``breathy,'' ``twangy,'' etc. This subjectivity, while capturing the nuanced perceptions of human listeners, presents challenges for consistent and accurate analysis, especially in applications requiring precise differentiation of voice characteristics. Such examples include speaker profiling for security and law enforcement, where the automated deduction of a multitude of speaker characteristics from voice is desired, but their relationships to voice have only previously been documented in terms of subjective assessments of the corresponding voice qualities. 

The need for objective measurement arises from the limitations of subjective evaluation, which can vary widely between listeners and lack the precision necessary for applications such as medical diagnostics, forensic analysis, and automated speech processing. Objective measures can provide a standardized way to evaluate voice quality, enabling consistent assessments across different contexts and applications.

The goal of this paper is to offer a path toward achieving this goal - to propose objective assessment of voice qualities. By analyzing the physical properties of voice signals, many prior studies have established correlations between these low-level signal features and the perceived qualities of the human voice. Such studies have identified and noted many specific signal characteristics -- such as frequency, amplitude, pitch, and other spectral characterizations -- that relate to the subjective qualities traditionally used to describe voice. Leveraging these correlations, it is possible to develop formulae or algorithms that quantitatively assess voice quality based on measurable signal properties.

Towards this goal of quantitative expression and evaluation of voice quality, we propose a clear 3-step procedure: 1) We choose 24 voice quality features that are widely used in different fields, especially in clinical literature. For each of these, we collate the results of prior scientific studies that have documented the statistical relationships of these voice qualities to members of a set of 25 low-level signal characterizations. 2) We translate these observations into a set of 24 formulae with linear terms and learnable parameters. The rationale for these is explained in a later section of this paper. 3) We experimentally evaluate the consistency and accuracy of these formulae by devising human assessment and data-driven experiments, for which we also translate the low-level feature computation to their neural counterpart algorithms. We show that the formulae are consistent and strongly follow the expected subjective trends of voice quality assessment by trained humans.

\section{Related Work}
Given the usage and significance of voice quality in various scenarios especially in diagnosing health conditions, speaker profiling etc, over the years multiple methodologies have been developed for measuring voice quality. These efforts include visual analysis, perceptual evaluation, aerodynamic measures, acoustic analysis, and self-evaluation by the individual \cite{dejonckere2001basic,boominathan2014multi}. Below we give examples of some of these efforts.

Auditory perceptual assessment is the most widely used methodology for measuring voice quality. Laver’s Voice Profile Analysis \cite{maidment1981john} proposed a phonetic classification system, using differences in the laryngeal and supra-laryngeal settings of the voice production systems for describing differences in voice qualities. Other methods include evaluation protocols like GRBAS \cite{hirano1986clinical} scale. It evaluates an individual on 5 voice qualities including roughness, breathiness, asthenia, strain and grade of the voice. Another example is RBH scale \cite{nawka1994auditive} which evaluates 3 voice qualities including roughness, breathiness, and hoarseness. CAPE-V \cite{kempster2009consensus} evaluates several voice qualities including loudness, diplophonia, vocal fry, falsetto, asthenia, aphonia, pitch instability, tremor, wet/gurgly, roughness, breathiness and strain. \cite{koutsogiannaki2014importance} proposes a metric based on the phase spectrum, demonstrating its effectiveness in detecting voice disorders by capturing irregularities in the glottal source. They test their methodology on private medical dataset and verify their ranking of voice quality against doctor's rankings. Some other methodologies analyse power spectra of a given speech signal over time, which reflect differences in voice qualities. 

Other proposed models evaluate sustained vowels by an individual. Such models include Acoustic Voice Quality Index (i.e., AVQI) \cite{maryn2010toward} and Cepstral Spectral Index of Dysphonia \cite{awan2009estimating}, widely used in clinical settings to identify vocal abnormalities. 

While existing methodologies contribute to voice quality assessment, they often suffer from limitations. Some require expert analysis, while others impose specific requirements on the individual, such as sustaining vowels. Additionally, these methods typically focus on a subset of voice qualities, neglecting a comprehensive evaluation. This necessitates the development of readily deployable, user-friendly approaches that capture the full spectrum of voice qualities without imposing stringent pre-conditions.

\section{Voice qualities, signal characteristics and rationale for selection}

In this paper we focus on 24 voice qualities (VQF). These are listed below. 
For detailed definition and description of each of these qualities, please refer to \cite{singh2019profiling}.

\NumTabs{2}

\begin{inparaenum}[1.]
\noindent\item Coveredness (Cov)
\tab \item Aphonicity (Aph)
\tab \item Biphonicity (Biph)
\tab \item Breathiness (Brea)
\tab \item Creakiness (Crea)
\tab \item Diplophonicity (Dip)
\tab \item Flutter (Flu)
\tab \item Glottalization (Glo)
\tab \item Hoarseness (Hoa)
\tab \item Roughness (Rou)
\tab \item Nasality (Nas)
\tab \item Jitter (Jit)
\tab \item Pressed (Pre)
\tab \item Pulsed (Pul)
\tab \item Resonant (Res)
\tab \item Shimmer (Shim)
\tab \item Strained (Stra)
\tab \item Strohbassness (Stro)
\tab \item Tremor (Tre)
\tab \item Twanginess (Twa)
\tab \item Ventricular (Ven)
\tab \item Wobble (Wob)
\tab \item Yawniness (Yaw)
\tab \item Loudness (Lou)
\end{inparaenum}


We use 25 Low Level speech Features (LLF), whose abbreviations are found in the Table below, but whose detailed definitions can be referred to \cite{singh2019profiling}.

\subsection{Correlating LLF to VQF}

\begin{table*}[tbh!]
\caption{Table shows the correlation of 24 Voice Quality features (columns) with the 25 low level signal features (rows). For reference the color palette can be found in Table \ref{tab:color_palette}.}
    \centering
    \begin{tabular}{@{\extracolsep{\fill}}|l|c|c|c|c|c|c|c|c|c|c|c|c|}
        \hline
        \textbf{Feat}$\downarrow$ \hfill \textbf{VQuality}$\rightarrow$ & Cov & Aph & Biph & Brea & Crea & Dip & Flu & Glo & Hoa & Rou & Nas & Jit \\ \hline
Loudness             & \nn & \pp & - & \wn \cite{shrivastav2003objective} & \wn \cite{eskenazi1990acoustic} & - & \pp & - & \nn \cite{pontes2002characteristics} & - & - & \wn \\ \hline
alphaRatio           & \pp & \nn & \wpos & \nn \cite{klich1982relationships} & \pp \cite{monson2014perceptual} & \nn & - & \pp & \nn & \wn & \pp \cite{madill2019impact} & \wn \\  \hline
hammarbergIndex      & \nn & \nn & - & \strn \cite{klich1982relationships} & \nn & \nn & \wn & \pp & \nn & \wn & \pp \cite{madill2019impact} & - \\  \hline
slope0-500           & \pp & \nn & \wn & \nn \cite{childers1991vocal} & \pp \cite{wright2019voice} & \wpos & - & \pp & \pp & \wpos & \pp & \pp \\  \hline
slope500-1500        & \nn & \pp & \wpos & \nn \cite{childers1991vocal} & \nn \cite{childers1991vocal} & - & - & \nn & - & \ic & \nn & \pp \\ \hline
spectralFlux         & \nn & \nn & \pp & \cite{klatt1990analysis} \strp & \pp & \pp & \wpos & \pp & \pp \cite{pontes2002characteristics} & \pp \cite{pontes2002characteristics} & \pp & \pp \\ \hline
mfcc1                & \pp & \nn & - & \ic & \nn \cite{wright2019voice} & \wpos & \ic & \nn & - & \nn & \pp & - \\ \hline
mfcc2                & \pp & \nn & - & \ic & \nn \cite{wright2019voice}& - & \ic & \nn  &- & \nn & \pp & - \\ \hline
mfcc3                &  \pp & \nn & - & \ic & \nn \cite{wright2019voice}& - & \ic & \nn & - & \nn & \pp & -\\ \hline
mfcc4                & \pp & \nn & - & \ic & \nn \cite{wright2019voice}& \wpos & \ic & \nn & -& \nn & \pp & -\\ \hline
F0semitoneFrom27.5Hz & \pp & \nn & \pp & \wn \cite{shrivastav2003objective} & \nn \cite{eskenazi1990acoustic} & \pp & \ic & \nn & \nn \cite{eskenazi1990acoustic} & \wn  & - & \ic\\ \hline
jitterLocal          & \wn & \pp & \nn & \pp \cite{shrivastav2003objective} & \pp \cite{pearsell2023effects} & \pp & - & \pp & \pp \cite{eskenazi1990acoustic} & \pp & - & \strp \\ \hline
shimmerLocaldB       & \wn & \pp & - & \pp \cite{shrivastav2003objective} & \pp \cite{pearsell2023effects} & \pp & - & \pp & \pp & \pp & - & \wpos\\ \hline
HNRdBACF             & \pp & \nn & - & \nn \cite{shrivastav2003objective} & \nn \cite{eskenazi1990acoustic} & \nn & \wpos & \nn & \nn \cite{yumoto1982harmonics} & \strn \cite{eskenazi1990acoustic} & \nn & \wn\\ \hline
logRelF0-H1-H2       & \pp & \nn & - & \nn \cite{shrivastav2003objective} & \nn \cite{kumar2011vowel} & \wpos & \nn & \wpos & \pp \cite{narasimhan2017spectral} & \wn & \nn & - \\ \hline
logRelF0-H1-A3       &  \nn & \nn & - & \nn \cite{shrivastav2003objective}  & \pp  & \wpos & \nn & \ic & \strn  & \wn & \nn & - \\ \hline
F1frequency          & \strp & \pp & - & \strp \cite{klatt1990analysis}  & \nn  & \nn & - & \wpos & \nn \cite{pontes2002characteristics} & \wn \cite{pontes2002characteristics} & \nn & -\\ \hline
F1bandwidth          & - & \nn & - & \pp \cite{klatt1990analysis} & \nn \cite{esposito2021examining} & \wpos & - & \wpos & \pp  & \wpos  & \nn & \pp\\ \hline
F1amplitudeLogRelF0  & \wn & \nn & - & \nn \cite{klatt1990analysis} & \wn & \wn & - & \wpos & \nn \cite{pontes2002characteristics} & \wpos & \pp & \pp\\ \hline
F2frequency          & \wn & - & - & \strn  & \pp \cite{esposito2021examining} & \nn & - & \ic & \wn \cite{pontes2002characteristics} & \wn \cite{pontes2002characteristics} & \nn & \ic \\ \hline
F2bandwidth          & \wn & - & - & \strp \cite{klatt1990analysis} & \pp  & \strp & - & \wn & \wpos  & \wpos  & \nn & \wpos \\ \hline
F2amplitudeLogRelF0  & \wn & \nn & - & \nn & \nn & \nn & - & \wpos & \wpos & \wn & \pp \cite{pruthi2007analysis}  & -\\ \hline
F3frequency          & \wn & \pp & - & \strn \cite{ishi2010analysis} & \wpos \cite{gobl1989preliminary} & \nn & \pp & \pp & \nn \cite{pontes2002characteristics} & \wpos  & \strn & -\\ \hline
F3bandwidth          & - & - & \pp & \pp \cite{ishi2010analysis} & \wn \cite{gobl1989preliminary} & \wpos & \wpos & \pp & \pp   & \wn \cite{pontes2002characteristics} & \nn & -\\ \hline
F3amplitudeLogRelF0  & - & - & \pp & \strn & \wn & \wpos & - & \pp & \strn \cite{pontes2002characteristics} & \wn & \pp & - \\ 
\hline
\end{tabular}
\medskip

\vspace{1em}

    \begin{tabular}{@{\extracolsep{\fill}}|l|c|c|c|c|c|c|c|c|c|c|c|c|}
    \hline
\textbf{Feat}$\downarrow$ \hfill \textbf{VQuality}$\rightarrow$ & Pre & Pul & Res & Shim & Stra & Stro & Tre & Twa & Ven & Wob & Yaw & Lou \\ \hline
Loudness             & \wpos & \nn & \wpos & \wn & \pp & - & \pp & \wpos & \wn & - & - & \strp \\ \hline
alphaRatio           & - & \nn &  & - & \nn & - & \pp & \pp & \nn & - & - & - \\  \hline
hammarbergIndex      & - & \strp &  & - & \strp & - & \pp & \pp & \nn & - & \nn & - \\  \hline
slope0-500           & \ic & \pp & \nn  & - & \pp & \strp & \nn & \pp & \nn & \pp & \pp & \pp \\  \hline
slope500-1500        & \ic & \nn & \pp & - & \pp & \wn & \nn & \wpos & \pp & \pp & - & \pp \\ \hline
spectralFlux         & \ic & \ic & - & - & \pp & \strp & \pp & \pp & \pp & \pp & \pp & \pp \\ \hline
mfcc1                & \ic & \wpos & \strp & - & - & \pp & \pp & \strp & \pp & \pp & \pp & \pp \\ \hline
mfcc2                & \ic & \wpos & \strp & - & - & \nn & \pp & \ic & \pp & - & \pp & \pp \\ \hline
mfcc3                & \ic & - & \strp & - & - & \pp & \pp & \ic & \pp & \pp & \pp & \pp\\ \hline
mfcc4                & \ic & - & \strp & - & - & \nn & \pp & \ic & \pp & - & \pp & \pp\\ \hline
F0semitoneFrom27.5Hz & \pp & \nn & - & - & - & \nn & \pp & \wpos & \nn & \pp & \pp & \pp \\ \hline
jitterLocal          & - & \wpos & - & \wpos & \wpos & \wpos & \pp & \nn & \nn & - & \nn & \ic \\ \hline
shimmerLocaldB       & - & \wpos & - & \strp & - & \wpos & \pp & \wpos & \nn & - & \nn & \ic\\ \hline
HNRdBACF             & \nn & \ic & \pp & \wn & \wn  & \nn & \nn & \pp & \nn & - & \pp & \pp\\ \hline
logRelF0-H1-H2       & \pp & - & \pp & \wn & \nn & \nn & \strn & \strp & - & \nn & \pp & -\\ \hline
logRelF0-H1-A3       & \pp & - & \pp & \wn & \nn & \nn & \nn & \strp & - & \nn & \pp & -\\ \hline
F1frequency          & \pp & \nn & \nn & \nn & \nn & \nn & \nn & \nn & \nn & \nn & \strn & \nn\\ \hline
F1bandwidth          & \nn & \wn & \nn & \wn & \nn & \strn & \pp & \nn & \nn & \nn & \pp & \pp\\ \hline
F1amplitudeLogRelF0  & \pp & \wn & \strp & \wn & \nn & \ic & \nn & \pp & \pp & \wn & \nn & \ic \\ \hline
F2frequency          & \pp & \ic & \nn & \wn & \wn & \nn & \ic & \wpos & \nn & \ic & \pp & \pp \\ \hline
F2bandwidth          & \nn & \ic & \strn & \wn & \pp & \strp & \ic & \strn & \nn & \ic & \pp & \pp \\ \hline
F2amplitudeLogRelF0  & \pp & - & \pp & \wn & \nn & \nn & - & \pp & \pp & \nn & \pp & \pp \\ \hline
F3frequency          & \pp & - & \strn & \wn & \nn & \nn & - & \pp & \nn & \wpos & \nn & \pp\\ \hline
F3bandwidth          & \nn & - & \nn & \wpos & \nn & \pp & \pp & \nn & \nn & \wpos & \pp & \pp \\ \hline
F3amplitudeLogRelF0  & \pp & - & \strp & \wn & \nn & \nn & \ic & \pp & \pp & \nn & \ic & - \\ \hline
    \end{tabular}
    \label{tab:corr_table}
\end{table*}

\begin{table}[tbh!]
    \centering
    \caption{Color Palette for the Correlations Table}
    \footnotesize
    \begin{tabular}{|l|c|l|}
    \toprule
    Correlation & Color & Weight \\ 
    \midrule
       Strong Negative (SN)  & \strn & 1 \\ \hline
       Negative (N) & \nn & 0.75 \\ \hline
       Weak Negative (WN) & \wn & 0.25 \\ \hline
       Neutral (N) & - & 0\\  \hline
       Weak Positive (WP) & \wpos & 0.25 \\ \hline
       Positive (P) & \pp & 0.75 \\ \hline
       Strong Positive (SP)  & \strp & 1 \\  \hline
       Inconclusive (IC) & \ic & 0 \\ 
       \bottomrule
    \end{tabular}
    \label{tab:color_palette}
\end{table}



Correlations for voice qualities are established based on extensive analysis of prior literature. Table \ref{tab:corr_table} shows the correlations between 24 voice quality and 25 low level speech features. 
To give an example of how a column is filled up in Table \ref{tab:corr_table}, lets look at the voice quality of `Breathiness'.
\cite{hillenbrand1994acoustic, hillenbrand1996acoustic, stranik2014acoustic, labuschagne2016perception} demonstrates that breathiness has a negative correlation with loudness, alphaRatio (the ratio of the summed energy from 50-1000 Hz and 1-5 kHz), hammerbergIndex (the ratio of the strongest energy peak in the 0-2 kHz region to the strongest peak in the 2-5 kHz region), a positive correlation with spectral flux (a measure of the change in the spectral content of a sound over time), has a positive correlation with F1, F2 and F3 bandwidth and \cite{prytz1976longtime} shows that breathiness has a negative correlation with F2 and F3 frequency.

The other columns in Table \ref{tab:corr_table} have been similarly completed based on a collated analysis of findings in prior literature. For example, a subset of findings we refer to in this work are: nasality \cite{chen1996acoustic, yoshida2000spectral}, tremor \cite{dromey2002influence, shao2010acoustic}, creakiness \cite{villegas2023psychoacoustic, hildebrand2016creaky, yuasa2010creaky}, hoarseness \cite{bowler1964fundamental}, pressed \cite{grillo2008evidence, eskenazi1990acoustic}, loudness \cite{yanushevskaya2013voice}, and many others. The Table below includes as many citations as the space allows. 

For the specific example of breathiness given above, we can construct the corresponding formula as follows: 


\begin{align*}
&Breathiness =  1/21 \times ( + (0.25 * (\mu_L  - v_L) / \sigma_L ) \\
             &+ (0.75 \times (\mu_{aR}  - v_{aR}) / \sigma_{aR} ) + (1 \times (\mu{hI}   - v_{hI}) / \sigma_{hI} )\\
             & + (0.75 \times (\mu_{s0-500}  - v_{s0-500}) / \sigma_{s0_500} ) \\
             & + (0.75 \times (\mu_{s500-1500}   - v_{s500-1500} ) / \sigma_{s500-1500}  )\\
             &+ (1 \times (v_{sF} - \mu_{sF}) / \sigma_{sF} )\\
             &+ (0.25 \times (\mu_{F0semi}   - v_{F0semi} ) / \sigma_{F0semi}  )\\
             &+ (0.75 \times (v_{jit} - \mu_{jit}) / \sigma_{jit} )\\
             &+ (0.75 \times (v_{shim} - \mu_{shim}) / \sigma_{shim} )\\
             &+ (0.75 \times (\mu_{hnr}  - v_{hnr}) / \sigma_{hnr} )\\
             &+ (0.75 \times (\mu_{lRH1H2}  - v_{lRH1H2}) / \sigma_{lRH1H2} )\\
             &+ (0.75 \times (\mu_{lRH1A3}  - v_{lRH1A3}) / \sigma_{lRH1A3} )\\
             &+ (1 \times (v_{F1f} - \mu_{F1f}) / \sigma_{F1f} )\\
             &+ (0.75 \times (v_{F1b} - \mu_{F1b}) / \sigma_{F1b} )+ (0.75 \times (\mu_{F1a}  - v_{F1a}) / \sigma_{F1a} )\\
             &+ (1 \times (\mu_{F2f}  - v_{F2f}) / \sigma_{F2f} )+ (1 \times (v_{F2b} - \mu_{F2b}) / \sigma_{F2b} )\\
             &+ (0.75 \times (\mu_{F2a}  - v_{F2a}) / \sigma_{F2a} )+ (1 \times (\mu_{F3f}  - v_{F3f}) / \sigma_{F3f} )\\
             &+ (0.75 \times (v_{F3b} - \mu_{F3b}) / \sigma_{F3b} )+ (1 \times (\mu_{F3a}  - v_{F3a}) / \sigma_{F3a} ) )
 \end{align*}

 The strategy given above can be generalized to obtain formulaic representations for other voice qualities. To do this we use the correlation information that we have collected in Table \ref{tab:corr_table}, as follows: A given voice quality $i$ ($vq_i$) can be written as a function of each of the low-level features $j$ ($\text{llf}_j$) as shown below:

\begin{equation}
    vq_i = 1/Z * \sum_j \frac{c_{i,j} * w_{i,j}  ( v_{\text{llf}_j} - \mu_{llf_j})}{\sigma_{\text{llf}_j}}
\end{equation}

where $Z$ is the normalizing factor which equals to the number of $\text{llf}$ involved in the equation. This ensures that the absolute value of the voice quality is between 0 and 1. $c_{i,j}$ is $1$ when the relation between $vq_i$ and $\text{llf}_j$ is a positive correlation, it is $-1$ if the correlation is negative, else it is 0. The $w_{i,j}$ is the weight assigned to the correlation. $\mu_{\text{llf}_j}$ and $\sigma_{\text{llf}_j}$ are the statistics of the $\text{llf}_j$ calculated as explained later \ref{subsubsec:extract_llf}. 

The complete formulae for all 24 voice qualities can be found in the following document: 
\url{https://github.com/ydhira/voicequeslityformulae/blob/master/VQ%20Formulas%20all.pdf}

\section{Experimental Validation}

\subsection{Methodology}

\subsubsection{Extracting Low level features} \label{subsubsec:extract_llf}
To extract the low level signal characteristics, we use the Temporal Acoustic Parameter (TAP) model \cite{zeng2023taploss}. The model is trained on speech data and is robust in predicting the low level features. 
To calculate the statistics (mean and variance) of the multiple low level speech features, we extract low level signal features from the NIST Speaker recognition evaluations (SRE) sets from year 2004, 2005, 2006 and 2008, with a total of 36612 utterances and 3805 speakers.

\subsection{Dataset}

\subsubsection{Expert labeled audios}
For each of the 24 voice quality, we search the web for any speech labeled for the voice quality. The speech samples were manually examined and verified. Only the relevant portion of the samples are spliced. We find data for 10 voice qualities, including breathiness, creakiness, roughness, resonant, which have more than 10 samples and for hoarseness, nasality, pressed, tremor, twanginess and yawniness which have lesser than 10 samples.  
The collected audio files can be found here:
\url{https://github.com/ydhira/voicequeslityformulae/tree/master}

Some voice qualities have a direct correlate to the low level speech features, like jitter, shimmer and loudness, notice that these three are both a low level feature and a voice quality. The difference between them can be explained by the following example: consider when a person is speaking softly and shouts for a short time, the signal level characteristic would denote this gradual decline and incline in the loudness feature, whereas perceptually, a listener might consider the speaker soft spoken, not loud. These three quantities are denoted with * in the following section. 



\subsection{Results}
\begin{table}[tbh!]
    \centering
    \caption{The Table presents the Voice Quality for which expert data is collected. The total number of pairs formed are shown in the next column, followed by the accuracy achieved by the established formulas in identifying the audio with the particular voice quality.}
    \begin{tabular}{llc}
    \toprule
        Voice Quality&Total Pairs &Acc (\%) \\
        &Pairs&\\ 
    \midrule
        Breathiness & 888  & 61.59 \\ 
        Creakiness &  1056 & 52.74 \\ 
        Hoarseness & 328 & 59.45 \\ 
        Roughness & 1000 & 60.02 \\ 
        Nasality  & 624 & 81.08 \\ 
        Jitter* & 6368 & 100.0 \\ 
        Pressed  & 85 & 98.82 \\ 
        Resonant & 825 & 53.94 \\ 
        Shimmer* & 6368 & 100.0 \\
        Tremor  & 168 & 78.57   \\
        Twanginess & 693 & 57.86 \\ 
        Yawniness & 693 & 39.11 \\ 
        Loudness* & 6368 & 62.20 \\ 
    \bottomrule
    \end{tabular}
    \label{tab:results_table}
\end{table}

For each voice quality, multiple pairs are formed where one sample comes from the dominant voice quality and the other comes from either a sample from another dominant voice quality or from neutral voice (voice that could not be assigned any of the voice qualities). The total numbers of such pairs formed are listed in the table~\ref{tab:results_table}. 

For a given voice quality $vq_i$, lets say a pair of speech samples is formed ($p_1, p_2$), where $p_1$ has the voice quality $vq_i$ dominantly, and $p_2$ doesnt have the voice quality, $vq_i$. We calculate the pair of scores ($s_1, s_2$) of the voice quality $vq_i$ using our presented formulas, where $s_1$ and $s_2$ are the scores of $p_1$ and $p_2$ respectively. If $s_1 > s_2$, we consider it a correct prediction, otherwise wrong. Table \ref{tab:results_table} present the accuracy achieved using the formulas for each of the 10 voice qualities for which data has been collected. 

We observe that voice quality denoted with * are the ones for which the low level feature is directly compared and not the voice quality. 
We observe that such voice qualities are performing the best, for example shimmer and jitter have an accuracy of 100\% over 6368 pair samples. Pressed, Nasality, Tremor also have high accuracy, all above 75\%. Yawniness has a low accuracy of 39.11\%. We believe this may be because it is a hard voice quality to differentiate from non-yawniness speech, and may also be because of the data that is collected, where some samples contain background music and may influence the results.
On average, the formulas are able to achieve an accuracy of 69.94\%. 

\section{Limitations and Conclusion}
There are several limitations to the current work. Firstly, the data collected for each voice quality is limited. To draw statistically significant conclusions, higher quality data would need to be collected. 
Our effort is small-scale, however it suffices to show the effectiveness and usefulness of the quantitative formulation of the voice qualities. 
Secondly, instead of just linear relationship between the low level speech signals and the voice qualities, more complex formulations could also be established. 

 
In conclusion, in this work we proposed formulation for extracting 24 different voice qualities from low level speech signal properties. This is the first work, to the best of our knowledge, which gives objective descriptions of voice qualities, rather than the traditional way of subjectively describing them. Since systematic data collection is non existent for different voice qualities, we crawl YouTube for speech samples for 10 voice qualities and manually clean the samples. We show how the proposed formulas perform on the collected data, achieving on average an accuracy of 69.64\%.
For future work, we aim to run a systematic analysis of the proposed formulas on large scale datasets, like Librispeech.

\newpage

\bibliographystyle{IEEEtran}
\bibliography{mybib}

\begin{thebibliography}{10}
\providecommand{\url}[1]{#1}
\csname url@samestyle\endcsname
\providecommand{\newblock}{\relax}
\providecommand{\bibinfo}[2]{#2}
\providecommand{\BIBentrySTDinterwordspacing}{\spaceskip=0pt\relax}
\providecommand{\BIBentryALTinterwordstretchfactor}{4}
\providecommand{\BIBentryALTinterwordspacing}{\spaceskip=\fontdimen2\font plus
\BIBentryALTinterwordstretchfactor\fontdimen3\font minus \fontdimen4\font\relax}
\providecommand{\BIBforeignlanguage}[2]{{%
\expandafter\ifx\csname l@#1\endcsname\relax
\typeout{** WARNING: IEEEtran.bst: No hyphenation pattern has been}%
\typeout{** loaded for the language `#1'. Using the pattern for}%
\typeout{** the default language instead.}%
\else
\language=\csname l@#1\endcsname
\fi
#2}}
\providecommand{\BIBdecl}{\relax}
\BIBdecl

\bibitem{sundberg1977acoustics}
J.~Sundberg, ``The acoustics of the singing voice,'' \emph{Scientific American}, vol. 236, no.~3, pp. 82--91, 1977.

\bibitem{boone2005voice}
D.~Boone, ``The voice and voice therapy,'' \emph{Allyn and Bacon google schola}, vol.~2, pp. 830--843, 2005.

\bibitem{klofstad2015perceptions}
C.~A. Klofstad, R.~C. Anderson, and S.~Nowicki, ``Perceptions of competence, strength, and age influence voters to select leaders with lower-pitched voices,'' \emph{PloS one}, vol.~10, no.~8, p. e0133779, 2015.

\bibitem{nolan1985phonetic}
F.~Nolan and H.~Hollien, ``The phonetic bases of speaker recognition by francis nolan,'' 1985.

\bibitem{dejonckere2001basic}
P.~H. Dejonckere, P.~Bradley, P.~Clemente, G.~Cornut, L.~Crevier-Buchman, G.~Friedrich, P.~Van De~Heyning, M.~Remacle, and V.~Woisard, ``A basic protocol for functional assessment of voice pathology, especially for investigating the efficacy of (phonosurgical) treatments and evaluating new assessment techniques: guideline elaborated by the committee on phoniatrics of the european laryngological society (els),'' \emph{European Archives of Oto-rhino-laryngology}, vol. 258, pp. 77--82, 2001.

\bibitem{boominathan2014multi}
P.~Boominathan, J.~Samuel, R.~Arunachalam, R.~Nagarajan, and S.~Mahalingam, ``Multi parametric voice assessment: Sri ramachandra university protocol,'' \emph{Indian Journal of Otolaryngology and Head \& Neck Surgery}, vol.~66, pp. 246--251, 2014.

\bibitem{maidment1981john}
J.~Maidment, ``The phonetic description of voice quality.'' \emph{Journal of the International Phonetic Association}, vol.~11, no.~2, pp. 78--84, 1981.

\bibitem{hirano1986clinical}
M.~Hirano and K.~R. McCormick, ``Clinical examination of voice by minoru hirano,'' 1986.

\bibitem{nawka1994auditive}
T.~Nawka, L.~C. Anders, and J.~Wendler, ``Die auditive beurteilung heiserer stimmen nach dem rbh-system,'' \emph{Sprache Stimme Geh{\"o}r}, vol.~18, no.~3, pp. 130--133, 1994.

\bibitem{kempster2009consensus}
G.~B. Kempster, B.~R. Gerratt, K.~V. Abbott, J.~Barkmeier-Kraemer, and R.~E. Hillman, ``Consensus auditory-perceptual evaluation of voice: development of a standardized clinical protocol,'' 2009.

\bibitem{koutsogiannaki2014importance}
M.~Koutsogiannaki, O.~Simantiraki, G.~Degottex, and Y.~Stylianou, ``The importance of phase on voice quality assessment,'' in \emph{Fifteenth Annual Conference of the International Speech Communication Association}, 2014.

\bibitem{maryn2010toward}
Y.~Maryn, P.~Corthals, P.~Van~Cauwenberge, N.~Roy, and M.~De~Bodt, ``Toward improved ecological validity in the acoustic measurement of overall voice quality: combining continuous speech and sustained vowels,'' \emph{Journal of voice}, vol.~24, no.~5, pp. 540--555, 2010.

\bibitem{awan2009estimating}
S.~N. Awan, N.~Roy, and C.~Dromey, ``Estimating dysphonia severity in continuous speech: application of a multi-parameter spectral/cepstral model,'' \emph{Clinical linguistics \& phonetics}, vol.~23, no.~11, pp. 825--841, 2009.

\bibitem{singh2019profiling}
R.~Singh, \emph{Profiling humans from their voice}.\hskip 1em plus 0.5em minus 0.4em\relax Springer, 2019, vol.~41.

\bibitem{shrivastav2003objective}
R.~Shrivastav and C.~M. Sapienza, ``Objective measures of breathy voice quality obtained using an auditory model,'' \emph{The Journal of the Acoustical Society of America}, vol. 114, no.~4, pp. 2217--2224, 2003.

\bibitem{eskenazi1990acoustic}
L.~Eskenazi, D.~G. Childers, and D.~M. Hicks, ``Acoustic correlates of vocal quality,'' \emph{Journal of Speech, Language, and Hearing Research}, vol.~33, no.~2, pp. 298--306, 1990.

\bibitem{pontes2002characteristics}
P.~A. Pontes, V.~P. Vieira, M.~I. Gon{\c{c}}alves, and A.~A. Pontes, ``Characteristics of hoarse, rough and normal voices: acoustic spectrographic comparative analysis,'' \emph{Rev Bras Otorrinolaringol}, vol.~68, no.~2, pp. 182--188, 2002.

\bibitem{klich1982relationships}
R.~J. Klich, ``Relationships of vowel characteristics to listener ratings of breathiness,'' \emph{Journal of Speech, Language, and Hearing Research}, vol.~25, no.~4, pp. 574--580, 1982.

\bibitem{monson2014perceptual}
B.~B. Monson, E.~J. Hunter, A.~J. Lotto, and B.~H. Story, ``The perceptual significance of high-frequency energy in the human voice,'' \emph{Frontiers in psychology}, vol.~5, p. 587, 2014.

\bibitem{madill2019impact}
C.~Madill, D.~D. Nguyen, K.~Yick-Ning~Cham, D.~Novakovic, and P.~McCabe, ``The impact of nasalance on cepstral peak prominence and harmonics-to-noise ratio,'' \emph{The Laryngoscope}, vol. 129, no.~8, pp. E299--E304, 2019.

\bibitem{childers1991vocal}
D.~G. Childers and C.~K. Lee, ``Vocal quality factors: Analysis, synthesis, and perception,'' \emph{the Journal of the Acoustical Society of America}, vol.~90, no.~5, pp. 2394--2410, 1991.

\bibitem{wright2019voice}
R.~Wright, C.~Mansfield, and L.~Panfili, ``Voice quality types and uses in north american english,'' \emph{Anglophonia. French Journal of English Linguistics}, no.~27, 2019.

\bibitem{klatt1990analysis}
D.~H. Klatt and L.~C. Klatt, ``Analysis, synthesis, and perception of voice quality variations among female and male talkers,'' \emph{the Journal of the Acoustical Society of America}, vol.~87, no.~2, pp. 820--857, 1990.

\bibitem{pearsell2023effects}
S.~Pearsell and D.~Pape, ``The effects of different voice qualities on the perceived personality of a speaker,'' \emph{Frontiers in Communication}, vol.~7, p. 909427, 2023.

\bibitem{yumoto1982harmonics}
E.~Yumoto, W.~J. Gould, and T.~Baer, ``Harmonics-to-noise ratio as an index of the degree of hoarseness,'' \emph{The journal of the Acoustical Society of America}, vol.~71, no.~6, pp. 1544--1550, 1982.

\bibitem{kumar2011vowel}
B.~R. Kumar, J.~S. Bhat, and P.~Mukhi, ``Vowel harmonic amplitude differences in persons with vocal nodules,'' \emph{Journal of Voice}, vol.~25, no.~5, pp. 559--561, 2011.

\bibitem{narasimhan2017spectral}
S.~Narasimhan and K.~Vishal, ``Spectral measures of hoarseness in persons with hyperfunctional voice disorder,'' \emph{Journal of Voice}, vol.~31, no.~1, pp. 57--61, 2017.

\bibitem{esposito2021examining}
C.~M. Esposito, M.~Sleeper, and K.~Sch{\"a}fer, ``Examining the relationship between vowel quality and voice quality,'' \emph{Journal of the International Phonetic Association}, vol.~51, no.~3, pp. 361--392, 2021.

\bibitem{pruthi2007analysis}
T.~Pruthi, \emph{Analysis, vocal-tract modeling and automatic detection of vowel nasalization}.\hskip 1em plus 0.5em minus 0.4em\relax University of Maryland, College Park, 2007.

\bibitem{ishi2010analysis}
C.~Ishi, H.~Ishiguro, and N.~Hagita, ``Analysis of the roles and the dynamics of breathy and whispery voice qualities in dialogue speech,'' \emph{EURASIP Journal on Audio, Speech, and Music Processing}, vol. 2010, pp. 1--12, 2010.

\bibitem{gobl1989preliminary}
C.~Gobl, ``A preliminary study of acoustic voice quality correlates,'' \emph{STL-QPSR}, vol.~4, no. 9-21, p. 534, 1989.

\bibitem{hillenbrand1994acoustic}
J.~Hillenbrand, R.~A. Cleveland, and R.~L. Erickson, ``Acoustic correlates of breathy vocal quality,'' \emph{Journal of Speech, Language, and Hearing Research}, vol.~37, no.~4, pp. 769--778, 1994.

\bibitem{hillenbrand1996acoustic}
J.~Hillenbrand and R.~A. Houde, ``Acoustic correlates of breathy vocal quality: Dysphonic voices and continuous speech,'' \emph{Journal of Speech, Language, and Hearing Research}, vol.~39, no.~2, pp. 311--321, 1996.

\bibitem{stranik2014acoustic}
A.~Str{\'a}n{\'\i}k, R.~{\v{C}}mejla, and J.~Vok{\v{r}}{\'a}l, ``Acoustic parameters for classification of breathiness in continuous speech according to the grbas scale,'' \emph{Journal of Voice}, vol.~28, no.~5, pp. 653--e9, 2014.

\bibitem{labuschagne2016perception}
I.~B. Labuschagne and V.~Ciocca, ``The perception of breathiness: Acoustic correlates and the influence of methodological factors,'' \emph{Acoustical Science and Technology}, vol.~37, no.~5, pp. 191--201, 2016.

\bibitem{prytz1976longtime}
S.~Prytz and B.~Fr{\"o}kjaer-Jensen, ``Longtime average spectra analyses of normal and pathological voices,'' \emph{Folia Phoniatr}, vol.~28, p. 280, 1976.

\bibitem{chen1996acoustic}
M.~Y. Chen, ``Acoustic correlates of nasality in speech,'' Ph.D. dissertation, Massachusetts Institute of Technology, 1996.

\bibitem{yoshida2000spectral}
H.~Yoshida, Y.~Furuya, K.~Shimodaira, T.~Kanazawa, R.~Kataoka, and K.~Takahashi, ``Spectral characteristics of hypernasality in maxillectomy patients 1,'' \emph{Journal of Oral Rehabilitation}, vol.~27, no.~8, pp. 723--730, 2000.

\bibitem{dromey2002influence}
C.~Dromey, P.~Warrick, and J.~Irish, ``The influence of pitch and loudness changes on the acoustics of vocal tremor,'' 2002.

\bibitem{shao2010acoustic}
J.~Shao, J.~K. MacCallum, Y.~Zhang, A.~Sprecher, and J.~J. Jiang, ``Acoustic analysis of the tremulous voice: assessing the utility of the correlation dimension and perturbation parameters,'' \emph{Journal of communication disorders}, vol.~43, no.~1, pp. 35--44, 2010.

\bibitem{villegas2023psychoacoustic}
J.~Villegas, S.~J. Lee, J.~Perkins, and K.~Markov, ``Psychoacoustic features explain creakiness classifications made by naive and non-naive listeners,'' \emph{Speech Communication}, vol. 147, pp. 74--81, 2023.

\bibitem{hildebrand2016creaky}
N.~Hildebrand-Edgar, ``Creaky voice: An interactional resource for indexing authority,'' Ph.D. dissertation, 2016.

\bibitem{yuasa2010creaky}
I.~P. Yuasa, ``Creaky voice: A new feminine voice quality for young urban-oriented upwardly mobile american women?'' \emph{American Speech}, vol.~85, no.~3, pp. 315--337, 2010.

\bibitem{bowler1964fundamental}
N.~W. Bowler, ``A fundamental frequency analysis of harsh vocal quality,'' \emph{Communications Monographs}, vol.~31, no.~2, pp. 128--134, 1964.

\bibitem{grillo2008evidence}
E.~U. Grillo and K.~Verdolini, ``Evidence for distinguishing pressed, normal, resonant, and breathy voice qualities by laryngeal resistance and vocal efficiency in vocally trained subjects,'' \emph{Journal of Voice}, vol.~22, no.~5, pp. 546--552, 2008.

\bibitem{yanushevskaya2013voice}
I.~Yanushevskaya, C.~Gobl, and A.~N{\'\i}~Chasaide, ``Voice quality in affect cueing: does loudness matter?'' \emph{Frontiers in psychology}, vol.~4, p. 52044, 2013.

\bibitem{zeng2023taploss}
Y.~Zeng, J.~Konan, S.~Han, D.~Bick, M.~Yang, A.~Kumar, S.~Watanabe, and B.~Raj, ``Taploss: A temporal acoustic parameter loss for speech enhancement,'' in \emph{ICASSP 2023-2023 IEEE International Conference on Acoustics, Speech and Signal Processing (ICASSP)}.\hskip 1em plus 0.5em minus 0.4em\relax IEEE, 2023, pp. 1--5.

\end{thebibliography}

\end{document}